\documentclass[12pt]{article}
\usepackage{latexsym}
\usepackage{amsmath,amssymb}
\usepackage{bm}
\usepackage[dvipdfmx]{graphicx}
\usepackage[dvipdfmx]{xcolor}
\usepackage{amssymb}
\usepackage{cite}
\usepackage{amsfonts, amscd, amsthm}
\usepackage[all]{xy}
\usepackage{here}
\usepackage{ulem}
\usepackage{mathabx}

\renewcommand{\theequation}{\thesection.\arabic{equation}}

\newcommand{\dd}{{\rm d}}
\newcommand{\MA}{{\mathcal A}}
\newcommand{\MB}{{\mathcal B}}
\newcommand{\MC}{{\mathcal C}}
\newcommand{\Z}{\mathbb{Z}}

\makeatletter
\@addtoreset{equation}{section}
\renewcommand{\theequation}{\thesection.\@arabic\c@equation}
\makeatother

\makeatletter
\renewcommand\appendix{\par
  \setcounter{section}{0}%
  \setcounter{subsection}{0}%
  \gdef\thesection{Appendix \@Alph\c@section }
  \renewcommand{\theequation}
  {\Alph{section}.\arabic{equation}}
}
\makeatother

\makeatletter
\newcounter{subeqncnt}
\def\thesubeqncnt{\alph{subeqncnt}}
\def\subequations{\begingroup%
\stepcounter{equation}\edef\@tempa{\theequation}%
\let\c@equation\c@subeqncnt\c@subeqncnt\z@
\edef\theequation{\@tempa\noexpand\thesubeqncnt}}

\makeatother

\allowdisplaybreaks

\begin{document}

\titlepage

\title{Two Flows 
Kowalevski Top as the Full Genus Two 
Jacobi's Inversion Problem and Sp(4,$\mathbb{R}$) Lie Group Structure} 
\author{Masahito Hayashi\thanks{masahito.hayashi@oit.ac.jp}\\
Osaka Institute of Technology, Osaka 535-8585, Japan\\
Kazuyasu Shigemoto\thanks{shigemot@tezukayama-u.ac.jp} \\
Tezukayama University, Nara 631-8501, Japan\\
Takuya Tsukioka\thanks{tsukioka@bukkyo-u.ac.jp}\\
Bukkyo University, Kyoto 603-8301, Japan\\
}
\date{\empty}


\maketitle
\abstract{ 
By using the first and second flows of the Kowalevski top, we can recreate the Kowalevski top into 
two-flows Kowalevski top, which has two-time variables. Then, we demonstrate that equations of 
the two-flows Kowalevski top become those of the full genus two Jacobi inversion problem. 
In addition to the Lax pair for the first flow, we construct a Lax pair for the second flow. 
Using the first and second flows, we demonstrate that the Lie group structure of these 
two Lax pairs is Sp(4,$\mathbb{R}$)/$\Z_2$ $\cong$ SO(3,2). With the two-flows Kowalevski 
top, we can conclude that the Lie group structure of the genus two hyperelliptic function is 
Sp(4,$\mathbb{R}$)/$\Z_2$ $\cong$ SO(3,2).  
}

\section{Introduction} 
\setcounter{equation}{0}

We are interested in the reason why exact solutions for some special non-linear differential 
equations exist, as well as a series of infinitely many solutions in some cases. 
Soliton equations are examples of such equations, such that various methods for studying 
soliton systems are beneficial to our objective. 
Starting from the inverse scattering method~\cite{Gardner,Lax,Zakhrov}, the soliton theory 
has several interesting 
developments, such as the AKNS formulation~\cite{Ablowitz}, geometrical 
approach~\cite{Bianchi,Hermann,Sasaki,Reyes}, B\"{a}cklund transformation~\cite{Wahlquist,Wadati1,Wadati2}, 
Hirota equation~\cite{Hirota1,Hirota2}, 
Sato theory~\cite{Sato},  vertex construction of the soliton 
solution~\cite{Miwa1,Date1,Jimbo1}, and Schwarzian type mKdV/KdV equation~\cite{Weiss}. 

We have a dogma that the reason why some special non-linear differential equations 
have a series of infinitely many solutions is that such  non-linear systems have the 
Lie group structure. Owing to an addition formula of the Lie group, we obtain 
a series of infinitely many solutions. 
As a representation of the addition formula of the Lie group, 
algebraic functions will emerge, 
such as trigonometric/elliptic/hyperelliptic functions, as the solutions of such special non-linear differential 
equations. A product of Lie group elements is given by an addition of exponential arguments of corresponding 
Lie algebra, and such multiplication formula is usually 
called an addition formula of the Lie group. 
In the KdV case, the B\"{a}cklund formula plays a role of an addition formula to 
provide new soliton solutions.

In our previous papers, using the Lie group structure 
SL(2,$\mathbb{R}$)/$\Z_2$ $\cong$ SO(2,1) $\cong$SU(1,1)/$\Z_2$ $\cong$ Sp(2,$\mathbb{R}$)/$\Z_2$ as 
the guiding principle, we revisited and studied why two dimensional integrable 
models, such as KdV/mKdV/sinh-Gordon, have quite optimal properties such as 
having a series of infinitely many solutions from the perspective of the Lie group's structure \cite{Hayashi1,Hayashi2,Hayashi3,Hayashi4,Hayashi5,Hayashi6}.

Here, we would like to revisit the famous Kowalevski top~\cite{Kowalevski, Kotter}. It is quite surprising that 
the Kowalevski top was first solved more than one hundred years ago, yet it remains in progress~\cite{Perelomov1,Reyman1,Haine,Adler,Reyman2,Marshall,Markushevich,Kuznetsov,Perelomov2,Marcelli,Dragovic}.
The original first flow (time-dependent) Kowalevski top can be formulated 
into the special genus two Jacobi's inversion problem in the form
\[
  \frac{\dd X_1}{\sqrt{f_5(X_1)}} + \frac{\dd X_2}{\sqrt{f_5(X_2)}}=0,
  \quad
  \frac{X_1 \dd X_1}{\sqrt{f_5(X_1)}}+\frac{X_2\dd X_2}{\sqrt{f_5(X_2)}}=i \dd t, 
\]
where $f_5(X)$ denotes some fifth-degree polynomial function of $X$.

In this research, we study the generalized two-flows Kowalevski top by using the 
second flow of the Kowalevski top~\cite{Adler,Haine}. We will demonstrate that 
equations of the two-flows Kowalevski top become those of the full genus two Jacobi's inversion problem. 
Next, in addition to the Lax pair of the original Kowalevski top~\cite{Reyman1,Reyman2}, we 
will provide the Lax pair for the second flow of the Kowalevski top. 
Using the Lax pairs of the first and second flows, we 
will demonstrate that Lax pairs have SO(3,2)$\cong$Sp(4,$\mathbb{R}$)/$\Z_2$ Lie 
group structure. Combining these two results, we conclude that the genus 
two hyperelliptic function 
has the SO(3,2)$\cong$Sp(4,$\mathbb{R}$)/$\Z_2$ Lie group structure.

\section{Two flows Kowalevski top as the full genus two Jacobi's inversion problem} 
\setcounter{equation}{0}

\subsection{Review of Kowalevski's work}
We first review Kowalevski's work~\cite{Kowalevski,Kotter}. A set of 
equations that describe the rotation of a rigid body 
around a fixed point under uniform gravitational force with a magnitude of  $Mg$ is given in the form
\begin{alignat*}{2}
A \frac{\dd \omega_1}{\dd t}&=(B-C) \omega_2 \omega_3
-Mg(\eta_0 \gamma_3-\zeta_0 \gamma_2), \qquad&\frac{\dd \gamma_1}{\dd t}=\omega_3 \gamma_2-\omega_2 \gamma_3,
\\
B\frac{\dd \omega_2}{\dd t}&=(C-A) \omega_3 \omega_1
-Mg(\zeta_0 \gamma_1 -\xi_0 \gamma_3), &\frac{\dd \gamma_2}{\dd t}=\omega_1 \gamma_3-\omega_3 \gamma_1,
\\
C\frac{\dd \omega_3}{\dd t}&=(A-B) \omega_1 \omega_2
-Mg(\xi_0 \gamma_2 -\eta_0 \gamma_1) , &\frac{\dd \gamma_3}{\dd t}=\omega_2 \gamma_1-\omega_1 \gamma_2.
\end{alignat*}
%
The coordinates of the center of mass and of the angular velocity are given in a frame attached to the 
rigid body with the fixed point as the origin, whose axes coincide with the principal 
axes of inertia. $(A, B, C)$ represent its principal moments of 
inertia, while $(\xi_0, \eta_0, \zeta_0)$ denotes the 
corresponding coordinates of the center of mass of 
the rigid body. Direction cosines $(\gamma_1,\gamma_2,\gamma_3)$ indicate 
the direction of the axis of rotation, 
while an angular velocity vector $(\omega_1,\omega_2,\omega_3)$ 
describes the rotation about the axis. The direction cosines are 
taken between the vertical axis (pointing down) and the three axes of the rigid body frame. 
This system has six variables.

For the Kowalevski top, we take $A=B=2C,~\zeta_0=0$, which provides  four 
conserved quantities. In this case, the center of mass is on 
the $\xi\eta$ plane. Owing to the fact that $A=B$, we can rotate 
the $\xi$ and $\eta$ axes around  $\zeta$ axis. Hence, 
we can put $\eta_0=0$ without losing generality. Then, the set of equations for the Kowalevski top becomes
\begin{align}
2\frac{\dd \omega_1}{\dd t}&=\omega_2 \omega_3 , 
\label{2e1}\\
2\frac{\dd \omega_2}{\dd t}&=-\omega_3 \omega_1-c_0\gamma_3  , 
\label{2e2}\\
\frac{\dd \omega_3}{\dd t}&=c_0 \gamma_2  ,
\label{2e3}\\
\frac{\dd \gamma_1}{\dd t}&=\omega_3 \gamma_2-\omega_2 \gamma_3, 
\label{2e4}\\
\frac{\dd \gamma_2}{\dd t}&=\omega_1 \gamma_3-\omega_3 \gamma_1 , 
\label{2e5}\\
\frac{\dd \gamma_3}{\dd t}&=\omega_2 \gamma_1-\omega_1 \gamma_2 . 
\label{2e6}
\end{align}
Here, we have used $c_0=-Mg \xi_0/C$ as a convenient notation.

In this Kowalevski's case, we define new variables as
\begin{align}
  \widehat{\gamma_1}&=\gamma_1 + \frac{\omega_1^2-\omega_2^2}{c_0},
  \\
  \widehat{\gamma_2}&=\gamma_2 + \frac{2\omega_1\omega_2}{c_0}.
\end{align}
These variables satisfy the following equations 
\begin{equation}
 \frac{\dd \widehat{\gamma_1}}{\dd t}=\omega_3 \widehat{\gamma_2},
 \quad
 \frac{\dd \widehat{\gamma_2}}{\dd t}=-\omega_3 \widehat{\gamma_1}.
\label{2e11}
\end{equation}
These equations demonstrate that $\widehat{\gamma_1}^2+\widehat{\gamma_2}^2$ is 
conserved, which 
is the Kowalevski's conserved quantity.

Now, we summarize four conserved quantities in the Kowalevski top.
\begin{align}
\textrm{a})&\quad \text{Energy:}\quad
2\omega_1^2+2\omega_2^2+\omega_3^2-2 c_0\gamma_1=6\ell_1
\label{2e12}\\
\textrm{b})&\quad \text{Vertical component of angular momentum:}\quad  
2\omega_1 \gamma_1+2\omega_2 \gamma_2+\omega_3 \gamma_3=2\ell\ 
\label{2e13}\\
\textrm{c})&\quad \text{Direction cosines:}\quad 
\gamma_1^2+\gamma_2^2+\gamma_3^2=1
\label{2e14}\\
\textrm{d})&\quad \text{Kowalevski's conserved quantity:}
\quad(c_0\widehat{\gamma_1})^2+(c_0\widehat{\gamma_2})^2=k^2
\label{2e15}
\end{align}
Here, we have introduced the parameters $\ell_1,~\ell$, and $k$ that represent values of each conserved 
quantity, which were used in the original Kowalevski's work~\cite{Kowalevski}. In the following, 
we adopt a convenient complex valuable defined as
\begin{equation}
  \xi 
  = 
  c_0\left(\widehat{\gamma_1} + i\widehat{\gamma_2}\right)
  =
  c_0(\gamma_1+i \gamma_2) + \left(\omega_1+i\omega_2\right)^2. 
\label{eq:def_of_xi}
\end{equation}
We regard $\xi$ and its complex conjugate $\bar{\xi}$ as 
independent valuables 
and denote $\xi_1$ and $\xi_2$, respectively. Note that $|\xi|^2=\xi_1\xi_2=k^2$. 
In addition, we introduce other complex variables $x_1$ and $x_2$ as
\begin{equation}
  x_1 = \omega_1+i\omega_2,\quad x_2 = \omega_1-i\omega_2.
\label{eq:def_of_x}
\end{equation}
We also consider $x_1$ and $x_2$ to be independent. In the following, 
we derive differential equations satisfied by $x_1$ and $x_2$, which will lead to the Jacobi's inversion problem.

By using these conserved quantities, $\omega_3$ and $\gamma_3$ are solved with variables of Eqs.(\ref{eq:def_of_xi}) 
and (\ref{eq:def_of_x}) in the form
\begin{align}
\omega_3^2&=6\ell_1-(x_1+x_2)^2+\xi_1+\xi_2=\MA+\xi_1+\xi_2 , 
\label{2e16}\\
c_0 \omega_3 \gamma_3&=2 \ell c_0+x_1 x_2 (x_1+x_2) -x_1 \xi_2-x_2 \xi_1
=\MB -x_1 \xi_2-x_2 \xi_1 , 
\label{2e17}\\
c_0^2 \gamma_3^2&=c_0^2-k^2-x_1^2 x_2^2+x_1^2 \xi_2+x_2^2 \xi_1
=\MC+x_1^2 \xi_2+x_2^2 \xi_1 , 
\label{2e18}
\end{align}
where we used 
$x_1^2 x_2^2=(\omega_1^2+\omega_2^2)^2$,
$x_1^2 \xi_2+ x_2^2 \xi_1=2(\omega_1^2+\omega_2^2)^2
+2 c_0  \left((\omega_1^2-\omega_2^2)\gamma_1
+2 \omega_1 \omega_2 \gamma_2\right)$. 
In the right-hand sides of Eqs.(\ref{2e16})--(\ref{2e18}), we have separated the expressions into 
two parts, i.e., a part which depends on $\xi_1, \xi_2$ and others with  
$$
\MA=6\ell_1-(x_1+x_2)^2,  
\qquad 
\MB=2 \ell c_0+x_1 x_2 (x_1+x_2), 
\qquad
\MC=c_0^2-k^2-x_1^2 x_2^2.
$$
From an identity 
$\omega_3^2 \times c_0^2 \gamma_3^2=(c_0 \omega_3 \gamma_3)^2$,
we have
\begin{align}
0&=(\MA+\xi_1+\xi_2) (\MC+x_1^2 \xi_2+x_2^2 \xi_1)-(\MB -x_1 \xi_2-x_2 \xi_1)^2
\nonumber\\
&=\MA \MC-\MB^2+k^2(x_1-x_2)^2+(\MA x_1^2+2 \MB x_1+\MC) \xi_2
+(\MA x_2^2+2 \MB x_2+\MC) \xi_1
\nonumber\\
&=R_1(x_1,x_2)+k^2 (x_1-x_2)^2+R(x_1) \xi_2+R(x_2) \xi_1, 
\label{2e19}
\end{align}
where
\begin{align}
R(x)&=\MA x^2+2 \MB x+\MC=-x^4 +6 \ell_1 x^2 +4 \ell c_0 x+c_0^2-k^2,
\label{2e20}\\
R_1(x_1, x_2)&=\MA\MC-\MB^2
\nonumber 
\\
&=-6 \ell_1 x_1^2 x_2^2-4 \ell c_0 (x_1+x_2)x_1 x_2-(c_0^2-k^2)(x_1+x_2)^2+6 \ell_1 (c_0^2-k^2)-4 \ell^2 c_0^2 .
\label{2e22}
\end{align}

To this point, we have adopted six dynamical 
variables $x_1, x_2, \xi_1, \xi_2, \omega_3, \gamma_3$ so far. 
Assuming that $x_1$ and $x_2$ have 
been solved, $\xi_1$ and $\xi_2$ can be obtained in principle from the relationship 
\begin{equation}
(\xi_1-x_1^2)\cdot(\xi_2-x_2^2)=c_0^2(\gamma_1^2+\gamma_2^2)=c_0^2(1-\gamma_3^2)
=c_0^2-(\MC+x_1^2 \xi_2+x_2^2 \xi_1), 
\label{2e23}
\end{equation}
and Eq.(\ref{2e19}). In addition, $\omega_3$ and $\gamma_3$ could be obtained from Eqs.(\ref{2e16}) 
and (\ref{2e18}), respectively. Accordingly, we can reduce the number of dynamical 
variables from six to two, i.e., $x_1$ and $x_2$ with four conserved quantities. In the calculation process  below, there appears the following 
combination $R(x_1) R(x_2)-(x_1-x_2)^2 R_1(x_1,x_2)$, which gives 
\begin{equation}
R(x_1) R(x_2)-(x_1-x_2)^2 R_1(x_1,x_2)=R(x_1,x_2)^2 , 
\label{2e24}
\end{equation}
where
\begin{align}
R(x_1,x_2)&=\MA x_1 x_2+\MB (x_1+x_2)+\MC \nonumber\\
&=-x_1^2 x_2^2 +6 \ell_1 x_1 x_2 +2 \ell c_0 (x_1+x_2)+c_0^2-k^2.
\label{2e25}
\end{align}
We note that $R(x_1,x_1)=R(x_1)$ and $R(x_2,x_2)=R(x_2)$.

Let us consider differential equations for two dynamical variables $x_1$ and $x_2$:  
\begin{align}
 2\frac{\dd x_1}{\dd t}=2\left(\frac{\dd \omega_1}{\dd t}
+i  \frac{\dd \omega_2}{\dd t} \right)
&=-i \left( \omega_3 (\omega_1+i \omega_2)+c_0 \gamma_3 \right) 
 =-i  (\omega_3 x_1+ c_0 \gamma_3), 
\label{2e26}\\
  2\frac{\dd x_2}{\dd t}=2\left(\frac{\dd \omega_1}{\dd t}
-i \frac{\dd \omega_2}{\dd t}\right)
&=i\left( \omega_3 (\omega_1-i  \omega_2)+c_0 \gamma_3 \right) 
 =i  (\omega_3 x_2+ c_0 \gamma_3).
\label{2e27}
\end{align}
By using  Eqs.(\ref{2e16})--(\ref{2e18}), we have
\begin{align}
-4\left(\frac{\dd x_1}{\dd t}\right)^2
&=\omega_3^2 x_1^2 + 2 c_0 \omega_3 \gamma_3 x_1 + c_0^2\gamma_3^2 
\notag\\&=(\MA+\xi_1+\xi_2) x_1^2 + 2(\MB -x_1 \xi_2-x_2 \xi_1) x_1 + (\MC +x_1^2 \xi_2+x_2^2 \xi_1)
\notag\\
&=(\MA x_1^2+2\MB x_1 +\MC) + (x_1-x_2)^2 \xi_1
\notag\\
&=R(x_1)+(x_1-x_2) ^2\xi_1  ,
\label{2e28}\\
-4\left(\frac{\dd x_2}{\dd t}\right)^2
&=\omega_3^2 x_2^2 + 2 c_0 \omega_3 \gamma_3 x_2 + c_0^2\gamma_3^2 
\notag\\
&=(\MA+\xi_1+\xi_2) x_2^2 + 2(\MB -x_1 \xi_2-x_2 \xi_1) x_2 + (\MC +x_1^2 \xi_2+x_2^2 \xi_1)
\notag\\
&=(\MA x_2^2+2\MB x_2 +\MC) + (x_1-x_2)^2 \xi_2
\notag\\
&=R(x_2) + (x_1-x_2) ^2\xi_2  ,
\label{2e29}\\
4\frac{\dd x_1}{\dd t} \frac{\dd x_2}{\dd t}
&=\omega_3^2 x_1 x_2 + c_0 \omega_3 \gamma_3 (x_1+x_1) + c_0^2\gamma_3^2
\notag\\
&=(\MA+\xi_1+\xi_2) x_1 x_2+(\MB -x_1 \xi_2-x_2 \xi_1) (x_1+x_2)
+\MC +x_1^2 \xi_2+x_2^2 \xi_1
\notag\\
&=\MA x_1 x_2+\MB (x_1+x_2)+\MC
\notag\\
&=R(x_1,x_2)  .
\label{2e30}
\end{align}
Then, we have
\begin{align}
-\frac{4}{R(x_1)}\left(\frac{\dd x_1}{\dd t}\right)^2&=1+\frac{(x_1-x_2)^2}{R(x_1)} \xi_1 , 
\label{2e31}\\
-\frac{4}{R(x_2)}\left(\frac{\dd x_2}{\dd t}\right)^2&=1+\frac{(x_1-x_2)^2}{R(x_2)} \xi_2 ,
\label{2e32}\\
\frac{8}{\sqrt{R(x_1) R(x_2)}}
\frac{\dd x_1}{\dd t} \frac{\dd x_2}{\dd t}
&=\frac{2 R(x_1,x_2)}{\sqrt{R(x_1) R(x_2)}} .
\label{2e33}
\end{align}
Owing to the identity of Eq.(\ref{2e19}), we can 
eliminate $\xi_1,  \xi_2$ by considering the following combination of Eqs.(\ref{2e31})--(\ref{2e33})
\begin{align}
&-\frac{4}{R(x_1)}\left(\frac{\dd x_1}{\dd t}\right)^2
-\frac{4}{R(x_2)}\left(\frac{\dd x_2}{\dd t}\right)^2
\mp \frac{8}{\sqrt{R(x_1) R(x_2)}} \frac{\dd x_1}{\dd t} \frac{\dd x_2}{\dd t} \nonumber\\
&=\frac{2 R(x_1) R(x_2) +(x_1-x_2)^2 (R(x_1)\xi_2+R(x_2) \xi_1) 
\mp 2 R(x_1,x_2)\sqrt{R(x_1) R(x_2)}}{R(x_1) R(x_2)}\nonumber\\
&=\frac{2 R(x_1) R(x_2) -(x_1-x_2)^2 (R_1(x_1,x_2)+k^2 (x_1-x_2)^2) 
\mp 2 R(x_1,x_2)\sqrt{R(x_1) R(x_2)}} {R(x_1)R(x_2)}\nonumber\\
&=\frac{R(x_1) R(x_2) +R(x_1,x_2)^2 
\mp 2 R(x_1,x_2)\sqrt{R(x_1) R(x_2)} - k^2 (x_1-x_2)^4}{R(x_1) R(x_2)}
\nonumber\\
&=
\frac{4(x_1-x_2)^4}{R(x_1) R(x_2)} 
\left(\left(\frac{R(x_1,x_2) \mp \sqrt{R(x_1) R(x_2)}}{2(x_1-x_2)^2}\right)^2-\frac{k^2}{4}\right) , 
\label{2e34}
\end{align}
where we have used Eq.(\ref{2e24}). Then, we have 
\begin{equation}
\left(\frac{\dd x_1/\dd t}{\sqrt{R(x_1)}} \pm
\frac{\dd x_2/\dd t}{\sqrt{R(x_2)}} \right)^2 
=
-\frac{(x_1-x_2)^4}{R(x_1) R(x_2)} 
\left(\left(\frac{R(x_1,x_2) \mp \sqrt{R(x_1) R(x_2)}}{2(x_1-x_2)^2}\right)^2-\frac{k^2}{4}\right). \label{2e35}
\end{equation}

Next, we introduce Kowalevski variables $s_1, s_2$ in the form
\begin{equation}
s_1=\frac{R(x_1,x_2) -\sqrt{R(x_1) R(x_2)}}{2(x_1-x_2)^2}, \quad
s_2=\frac{R(x_1,x_2) +\sqrt{R(x_1) R(x_2)}}{2(x_1-x_2)^2}  . 
\label{2e36}
\end{equation}
In terms of the Kowalevski variables, we have 
\begin{equation}
\frac{\sqrt{R(x_1) R(x_2)}}{(x_1-x_2)^2}=s_2-s_1,
\label{2e37}
\end{equation}
while Eq.(\ref{2e35}) gives 
\begin{align}
\left(\frac{\dd x_1/\dd t}{\sqrt{R(x_1)}} +\frac{\dd x_2/\dd t}{\sqrt{R(x_2)}} \right)^2 
&=-\frac{1}{(s_2-s_1)^2} \left( s_1^2-\frac{k^2}{4}\right)  ,
\label{2e38}\\
\left(\frac{\dd x_1/\dd t}{\sqrt{R(x_1)}} -\frac{\dd x_2/\dd t}{\sqrt{R(x_2)}} \right)^2 
&=-\frac{1}{(s_2-s_1)^2} \left( s_2^2-\frac{k^2}{4}\right)  .
\label{2e39}
\end{align}
We now assume that the following $\varphi(s)$ exists
\begin{align}
\frac{\dd x_1/\dd t}{\sqrt{R(x_1)}} +\frac{\dd x_2/\dd t}{\sqrt{R(x_2)}} 
&=i \frac{\sqrt{s_1^2-k^2/4}}{s_1-s_2}
=\frac{\dd s_1/\dd t}{\sqrt{\varphi(s_1)}}  ,
\label{2e42}\\
-\frac{\dd x_1/\dd t}{\sqrt{R(x_1)}} +\frac{\dd x_2/\dd t}{\sqrt{R(x_2)}}  
&=i \frac{\sqrt{s_2^2-k^2/4}}{s_2-s_1} 
=\frac{\dd s_2/\dd t}{\sqrt{\varphi(s_2)}}   . 
\label{2e43}
\end{align}
Later, we will explicitly prove that $\varphi(s)$ is the third-degree polynomial 
function of $s$, which implies that the above assumption is 
true. Considering the square root of Eqs.(\ref{2e38}) 
and (\ref{2e39}), various signs emerge. 
After obtaining the explicit form of $\varphi(s)$, we return to Eqs.(\ref{2e42}) 
and (\ref{2e43}) ; in addition, via numerical calculations, we have 
checked that the above signs are correct. From the second and  
third terms of Eqs.(\ref{2e42}) 
and (\ref{2e43}), we have
\begin{equation}
\frac{\dd s_1}{\sqrt{f_5(s_1)} }
=i \frac{\dd t}{s_1-s_2}, \quad
\frac{\dd s_2}{\sqrt{f_5(s_2)} }
=-i \frac{\dd t}{s_1-s_2}  , 
\label{2e44}
\end{equation}
where
\begin{equation}
f_5(s)=\varphi(s) (s^2-k^2/4)   ,
\label{2e45}
\end{equation}
which gives a special genus two Jacobi's inversion problem of the form 
\begin{equation}
\frac{\dd s_1}{\sqrt{f_5(s_1)} }+\frac{\dd s_2}{\sqrt{f_5(s_2)} }
=0, \quad
\frac{s_1 \dd s_1}{\sqrt{f_5(s_1)} }+\frac{s_2 \dd s_2}{\sqrt{f_5(s_2)} }
=i \dd t  .
\label{2e46}
\end{equation}

A general genus two Jacobi's inversion problem is written in the form~\cite{Baker}
\begin{equation}
\dd u_1=\frac{\dd X_1}{Y_1}+\frac{\dd X_2}{Y_2}, \quad \dd u_2=\frac{X_1 \dd X_1}{Y_1}+\frac{X_2\dd X_2}{Y_2}, 
\quad Y=\sqrt{f_5(X)}   , 
\label{2e47}
\end{equation}
where $f_5(X)$ is a fifth-degree polynomial function of $X$. 
We will use the term ``full" 
Jacobi's inversion problem for the case where $\dd u_1 \ne 0$ and $\dd u_2 \ne 0$ 
and ``special" for $\dd u_1 = 0$ and $\dd u_2 \ne 0$. From Eq.(\ref{2e47}), we have
\begin{alignat}{2}
\frac{\partial X_1}{\partial u_2}&=\frac{Y_1}{X_1-X_2}, &\quad
\frac{\partial X_2}{\partial u_2}&=-\frac{Y_2}{X_1-X_2}  ,
\label{2e48}\\
\frac{\partial X_1}{\partial u_1}&=-\frac{X_2 Y_1}{X_1-X_2},  &\quad
\frac{\partial X_2}{\partial u_1}&=\frac{X_1 Y_2}{X_1-X_2}  .
\label{2e49}
\end{alignat}
By making the correspondence of $s_1 \leftrightarrow X_1$, $s_2 \leftrightarrow X_2$, $0 \leftrightarrow \dd u_1$, $i \dd t \leftrightarrow \dd u_2$, Eq.(\ref{2e47}) gives the special genus two Jacobi's inversion problem Eq.(\ref{2e46}).

If we notice that  Eq.(\ref{2e36}) is the change of variables $x_1, x_2 \rightarrow s_1(x_1,x_2), s_2(x_1,x_2)$, we 
can find $\varphi(s)$ in Eqs.(\ref{2e42}) and (\ref{2e43}) from the following dynamics 
independent identical relationships:
\begin{equation}
\frac{\dd x_1}{\sqrt{R(x_1)}} +\frac{\dd x_2}{\sqrt{R(x_2)}} 
=\frac{\dd s_1}{\sqrt{\varphi(s_1)}}, \quad
-\frac{\dd x_1}{\sqrt{R(x_1)}} +\frac{\dd x_2}{\sqrt{R(x_2)}}  
=\frac{\dd s_2}{\sqrt{\varphi(s_2)}}. 
\label{2e50}
\end{equation}
From Eqs.(\ref{2e50}), we have
\begin{equation}
\frac{\partial s_1}{\partial x_1}= \sqrt{ \frac{\varphi(s_1)}{R(x_1)}}, \quad
\frac{\partial s_1}{\partial x_2}= \sqrt{ \frac{\varphi(s_1)}{R(x_2)}}, \quad
\frac{\partial s_2}{\partial x_1}=-\sqrt{ \frac{\varphi(s_2)}{R(x_1)}}, \quad
\frac{\partial s_2}{\partial x_2}= \sqrt{ \frac{\varphi(s_2)}{R(x_2)}}. 
\label{eq:determine_phi}
\end{equation}
The existence of $\varphi(s)$ that satisfies Eqs.(\ref{eq:determine_phi}) 
implies that the 
original Kowalevski top can be expressed in the genus two Jacobi's inversion problem.

First, we consider the limit $x_2 \rightarrow x_1$ and take the most singular term. 
In this limit, $R(x_1,x_2) \rightarrow R(x_1)$, $R(x_2) \rightarrow R(x_1)$. Then, we have 
\begin{equation}
  s_2 \sim \frac{R(x_1)}{(x_1-x_2)^2},\quad
  \frac{\partial s_2}{\partial x_1} \sim -\frac{2R(x_1)}{(x_1-x_2)^3},\quad
  \frac{\partial s_2}{\partial x_2} \sim  \frac{2R(x_1)}{(x_1-x_2)^3},
\label{2e52}
\end{equation}
from Eq.(\ref{2e36}). By adopting Eqs.(\ref{eq:determine_phi}) and (\ref{2e52}), we obtain
\begin{equation}
  \frac{\partial s_2}{\partial x_1}\frac{\partial s_2}{\partial x_2}
  \sim -\frac{\varphi(s_2)}{R(x_1)}  
  \sim -\frac{4R(x_1)^2}{(x_1-x_2)^6}
  \sim -\frac{4s_2^3}{R(x_1)}.
\label{2e53}
\end{equation}
This implies that $\displaystyle{ \varphi(s_2) =4s_2^3+\text{(lower order)}}$, which is the 
third-degree polynomial function of $s$, i.e.,
\begin{equation}
  \varphi(s)=4s^3+k_2 s^2+k_1 s +k_0.
\label{2e54}
\end{equation}
The coefficients $k_2$, $k_1$, and $k_0$ are determined by using the relationship
\begin{equation}
\varphi(s_1)+\varphi(s_2)=\sqrt{R(x_1) R(x_2)} 
\left( \frac{\partial s_1}{\partial x_1} \frac{\partial s_1}{\partial x_2}
 - \frac{\partial s_2}{\partial x_1} \frac{\partial s_2}{\partial x_2} \right).
\label{2e55}
\end{equation}
$(x_1-x_2)^6 \times$(l.h.s of Eq.(\ref{2e55})) represents the polynomial 
of $x_1$ and $x_2$ that involves $k_2$, $k_1$, and $k_0$. By using the 
definitions of $s_1$ and $s_2$ given in Eqs.(\ref{2e36}), $(x_1-x_2)^6 \times$(r.h.s of Eq.(\ref{2e55})) 
is proven to be the polynomial of $x_1$ and $x_2$. Hence, 
from Eq.(\ref{2e55}), the coefficients $k_2$, $k_1$, and $k_0$ are determined. 
{Accordingly, we obtained $\varphi(s)$ in the form
\begin{equation}
\varphi(s)=4 s^3+6 \ell_1 s^2-(k^2-c_0^2) s -\frac{3}{2} \ell_1 (k^2-c_0^2)-\ell^2 c_0^2  . 
\label{2e56}
\end{equation}
By using Eq.(\ref{2e56}), we have checked Eqs.(\ref{eq:determine_phi}) to ensure 
that it is actually satisfied. We conclude that the original Kowalevski top can be 
expressed in the special genus two Jacobi's inversion problem Eq.(\ref{2e46}).

\subsection{Second flow of the Kowalevski Top}
We generalize the original Kowalevski top to two-flows Kowalevski top, and we replace 
$\omega_i (t) \rightarrow \omega_i (r, t)$, $\gamma_i( t) \rightarrow \gamma_i (r, t)$\ $(i=1,2,3)$ 
by introducing another time $r$, such that 
$x_1(t) \rightarrow x_1 (r, t)$, $x_2(t) \rightarrow x_2 (r, t)$, $s_1(t) \rightarrow s_1 (r, t)$, 
$s_2(t) \rightarrow s_2 (r, t)$ and $\dd/\dd t \rightarrow \partial/\partial t$. 
We call the set of Eqs.(\ref{2e1})--(\ref{2e6}) as the first flow equations. 
The second flow equations are given in the form~\cite{Haine,Adler} 
\begin{align}
2\frac{\partial \omega_1}{\partial r}&=-c_0 \gamma_2 \gamma_3 
+\omega_3 (\gamma_1 \omega_2-\gamma_2 \omega_1)
-2 \gamma_3 \omega_1 \omega_2
-\frac{1}{c_0} (\omega_1^2+\omega_2^2) \omega_2 \omega_3  , 
\label{2e57}\\
2\frac{\partial \omega_2}{\partial r}&=c_0 \gamma_1 \gamma_3 
+\omega_3 (\gamma_1 \omega_1+\gamma_2 \omega_2)
+\gamma_3(\omega_1^2- \omega_2^2)
+\frac{1}{c_0} (\omega_1^2+\omega_2^2) \omega_1 \omega_3  , 
\label{2e58}\\
\frac{\partial \omega_3}{\partial r}&=\gamma_2(\omega_1^2- \omega_2^2)
-2 \gamma_1 \omega_1 \omega_2 , 
\label{2e59}\\
\frac{\partial \gamma_1}{\partial r}&=\gamma_3
\left(\gamma_1 \omega_2-\gamma_2 \omega_1
-\frac{1}{c_0} (\omega_1^2+\omega_2^2) \omega_2\right)  , 
\label{2e60}\\
\frac{\partial \gamma_2}{\partial r}&=\gamma_3 
\left(\gamma_1 \omega_1+\gamma_2 \omega_2
+\frac{1}{c_0} (\omega_1^2+\omega_2^2) \omega_1\right)  , 
\label{2e61}\\
\frac{\partial \gamma_3}{\partial r}&=-\omega_2(\gamma_1^2+\gamma_2^2)
+\frac{1}{c_0}(\gamma_1 \omega_2-\gamma_2 \omega_1)(\omega_1^2+\omega_2^2) .   
\label{2e62}
\end{align}
Under these second flow equations, Eqs.(\ref{2e12})--(\ref{2e15}) are also conserved. 
Furthermore, the first and second flows are integrable, i.e., the integrability 
conditions of $\omega_i, \gamma_i$\ $(i=1,2,3)$ 
are satisfied. For example, the integrability condition 
$\partial_ r \partial_t \omega_1= \partial_t \partial_ r \omega_1$ is equivalent to   
\begin{equation}
\partial_r (\omega_2 \omega_3)
=\partial_t \left(-c_0 \gamma_2 \gamma_3 
+\omega_3 (\gamma_1 \omega_2-\gamma_2 \omega_1)
-2 \gamma_3 \omega_1 \omega_2
-\frac{1}{c_0} (\omega_1^2+\omega_2^2) \omega_2 \omega_3 \right).
\label{2e63}
\end{equation}
It can be observed that Eq.(\ref{2e63}) holds by using the first flow 
equations Eqs.(\ref{2e1})--(\ref{2e6}) and the second flow equations Eqs.(\ref{2e57})--(\ref{2e62}).

Let us denote
\begin{align}
\wp_{22}&=s_1+s_2=\frac{R(x_1, x_2)}{(x_1-x_2)^2}  , 
\label{2e64}\\
\wp_{21}&=-s_1 s_2=\frac{R(x_1)R(x_2)-R(x_1,x_2)^2 }{4(x_1-x_2)^4}
=\frac{R_1(x_1,x_2)}{4(x_1-x_2)^2}.
\label{2e65}
\end{align}
By using the first and second flow equations, we have checked
\[
\displaystyle{\frac{c_0}{2}\dfrac{\partial \wp_{22}}{\partial r}=\dfrac{\partial \wp_{21}}{\partial t}}
,\quad\text{i.e.},\quad 
\displaystyle{ \frac{c_0}{2}\frac{\partial}{\partial r} (s_1+s_2) =-\frac{\partial}{\partial t} (s_1 s_2)}.
\]
It can be considered as the integrability condition by identifying  $\wp_{22}, ~\wp_{21}$ as the genus 
two hyperelliptic $\wp$ functions with $\dd u_1 :\dd u_2= 2 \dd r /c_0 : \dd t$. This fact suggests 
that we have a full Jacobi inversion problem in the form
\begin{align}
\frac{\dd s_1}{\sqrt{f_5(s_1)}} +\frac{\dd s_2}{\sqrt{f_5(s_2)}}
&=\dd u_1=i \frac{2}{c_0} \dd r ,
\label{2e66}\\
\frac{s_1 \dd s_1}{\sqrt{f_5(s_1)}} +\frac{ s_2 \dd s_2}{\sqrt{f_5(s_2)}}
&=\dd u_2=i \dd t   .
\label{2e67}
\end{align}
Comparing Eqs.(\ref{2e66}) and (\ref{2e67}) with Eq.(\ref{2e47}), we have achieved the 
correspondence \mbox{$s_1 \leftrightarrow X_1$}, $s_2 \leftrightarrow X_2$,  
$\sqrt{f_5(s_1)} \leftrightarrow Y_1$, $\sqrt{f_5(s_2)} \leftrightarrow Y_2$, $\dd u_1 \leftrightarrow 2 i\dd r/c_0$, 
$\dd u_2 \leftrightarrow i \dd t$. Then, from Eqs.(\ref{2e48}) and (\ref{2e49}), we have 
\begin{alignat}{2}
\frac{1}{i} \frac{\partial s_1}{\partial t}&=\frac{ \sqrt{f_5(s_1)}}{s_1-s_2}, &\quad 
\frac{1}{i} \frac{\partial s_2}{\partial t}&=-\frac{ \sqrt{f_5(s_2)}}{s_1-s_2}  , 
\label{2e68}\\
\frac{c_0}{2 i} \frac{\partial s_1}{\partial r}&=-\frac{ s_2 \sqrt{f_5(s_1)}}{s_1-s_2} , &\quad
\frac{c_0}{2 i} \frac{\partial s_2}{\partial r}&=\frac{ s_1 \sqrt{f_5(s_2)}}{s_1-s_2}  . 
\label{2e69}
\end{alignat}
If two-flows Kowalevski top can be written in a full genus two Jacobi's inversion problem, Eqs.(\ref{2e68}) 
and (\ref{2e69}) must be satisfied. From Eq.(\ref{2e44}), it can be observed 
that Eq.(\ref{2e68}) is definitely satisfied. Then, we will check to confirm that 
Eq.(\ref{2e69}) is completely satisfied. The relative sign of Eqs.(\ref{2e68}) 
and (\ref{2e69}) is determined in the above value from the integrability conditions of $\wp_{22}$ and $\wp_{21}$. 
Then, we take the square and check the relationship. 
The first term of Eq.(\ref{2e69}) gives
\begin{align}
&-\frac{c_0^2}{4} \left(\frac{\partial s_1}{\partial r}\right)^2=\frac{s_2^2 f_5(s_1)}{(s_1-s_2)^2}
=\frac{s_2^2 \varphi(s_1)(s_1^2-k^2/4)}{(s_1-s_2)^2}\nonumber\\
&\Leftrightarrow \frac{(\partial s_1/\partial r)^2}{\varphi(s_1)}
=-\frac{4}{c_0^2} \frac{s_2^2 (s_1^2-k^2/4)}{(s_1-s_2)^2}   .  
\label{2e70}
\end{align}
The second term of Eq.(\ref{2e69}) gives
\begin{align}
&-\frac{c_0^2}{4} \left(\frac{\partial s_2}{\partial r}\right)^2=\frac{s_1^2 f_5(s_2)}{(s_1-s_2)^2}
=\frac{s_1^2 \varphi(s_2)(s_2^2-k^2/4)}{(s_1-s_2)^2}\nonumber\\
&\Leftrightarrow \frac{(\partial s_2/\partial r)^2}{\varphi(s_2)}
=-\frac{4}{c_0^2} \frac{s_1^2 (s_2^2-k^2/4)}{(s_1-s_2)^2}   .
\label{2e71}
\end{align}
However, by using Eq.(\ref{2e50}), we have
\begin{align}
\left( \frac{\partial x_1/\partial r}{\sqrt{R(x_1)}} +\frac{\partial x_2/\partial r}{\sqrt{R(x_2)}} \right)^2
&= \frac{(\partial s_1/\partial r)^2}{\varphi(s_1)}
=-\frac{4}{c_0^2} \frac{s_2^2 (s_1^2-k^2/4)}{(s_1-s_2)^2}  ,
\label{2e72}\\
\left( -\frac{\partial x_1/\partial r}{\sqrt{R(x_1)}} +\frac{\partial x_2/\partial r}{\sqrt{R(x_2)}} \right)^2
&=\frac{(\partial s_2/\partial r)^2}{\varphi(s_2)}
=-\frac{4}{c_0^2} \frac{s_1^2 (s_2^2-k^2/4)}{(s_1-s_2)^2}   .
\label{2e73}
\end{align}
The following equations are derived from the sum and difference between Eqs.(\ref{2e72}) and (\ref{2e73}); 
\begin{align}
\frac{(\partial x_1/\partial r)^2 }{R(x_1)}+\frac{(\partial x_2/\partial r)^2}{R(x_2)}
&=-\frac{1}{c_0^2} \frac{\left( 4(s_1 s_2)^2-k^2\left((s_1+s_2)^2-2 s_1 s_2\right)/2\right)}{(s_1-s_2)^2}
\nonumber\\
&=-\frac{1}{4 c_0^2} \frac{R_1(x_1,x_2)^2}{R(x_1)R(x_2)}
+\frac{k^2}{2 c_0^2} \frac{ R(x_1,x_2)^2+(x_1-x_2)^2 R_1(x_1,x_2)/2}{R(x_1) R(x_2)} , 
\label{2e75}\\
\frac{\partial x_1}{\partial r} \frac{\partial x_2}{\partial r}
&=\frac{k^2}{4 c_0^2} \frac{s_2+s_1}{s_2-s_1}\sqrt{R(x_1) R(x_2)}=\frac{k^2}{4 c_0^2} R(x_1,x_2) . 
\label{2e76}
\end{align}
Using Eqs.(\ref{2e57})--(\ref{2e62}), we have verified that Eqs.(\ref{2e75}) 
and (\ref{2e76}) are completely satisfied. 
Then, we can conclude that the first and second flows, which 
constitute the integrable two-flows Kowalevski top, provide the full genus two Jacobi inversion 
problem of  Eqs.(\ref{2e66}) and (\ref{2e67}).

\section{Lax pairs for two flows Kowalevski top and Sp(4,$\mathbb{R}$) Lie group structure} 
\setcounter{equation}{0}
We usually formulate integrable models with a Lax pair to explicitly 
demonstrate an integrability of models and systematically obtain various conserved 
quantities. Let us consider the KdV equation of the form $u_t -u_{xxx}+6u u_x=0$. 
From one Lax pair $L, B$ of the form 
\begin{equation}
L=-\partial_x^2+u, \quad  B=4 \partial_x^3-6 u \partial_x -3 u_x ,
\label{3e1}  
\end{equation}
%
$\partial_t L=[B, L]$ gives the KdV equation. From another Lax pair $L, B$, which is known as the 
AKNS formalism, of the form 
\begin{equation}
L=\left( \begin{array}{cc} \lambda & -u \\ -1  & -\lambda \end{array}\right), \quad 
B=\left( \begin{array}{cc}   
4 \lambda^3-2 \lambda u - u_x  & -u_{xx} -2 \lambda u_x -4 \lambda^2 u+2 u^2\\
-4 \lambda^2+2 u  & -4 \lambda^3 +2 \lambda u+u_x \end{array}\right) ,
\label{3e2}  
\end{equation}
where $\lambda$ represents a constant spectral parameter, 
$[\partial_x-L, \partial_t-B]=\partial_t L-\partial_x B-[B,L]=0$ gives the KdV equation. 
We construct a  Lax pair to examine 
the Lie group structure of an integrable model. 
As an infinitesimal Lie group transformation of an integrable model, we formulate a Lax 
pair using Lie algebra elements with only linear partial differential operators. 
The Lax pair of Eq.(\ref{3e2}) is appropriate for this purpose because it contains only the linear 
differential operators as $\partial_x$ and $\partial_t$. The matrices $L$ and $B$ are 
constructed from the \mbox{SL(2,$\mathbb{R}$)/$\Z_2$}$\cong$ SO(2,1) Lie algebra elements  
$\left( \begin{array}{@{}cc@{}} 1 & 0 \\ 0  & -1 \end{array}\right)$, 
$\left( \begin{array}{@{}cc@{}} 0 & 1 \\ 0  &  0 \end{array}\right)$, 
$\left( \begin{array}{@{}cc@{}} 0 & 0 \\ 1  &  0 \end{array}\right)$. 
Then, we can infer that the KdV equation has the 
SL(2,$\mathbb{R}$)/$\Z_2$ $\cong$ SO(2,1) Lie group structure.
 
In this study, we construct the Lax pairs of the first and 
second flows for two-flows Kowalevski top, to examine 
its Lie group structure. Reyman {\it et al.} have constructed the Lax pair 
for the first flow~\cite{Reyman1,Reyman2} by using the $4\times4$ spinor representation 
of SO(3,2) Lie algebra. Because SO(3,2) $\cong$ Sp(4,$\mathbb{R}$)/$\Z_2$, 
we formulate Lax pairs with the Sp(4,$\mathbb{R}$) Lie algebra.

\subsection{The bases for the Sp(4,$\mathbb{R}$) Lie algebra}
We first construct the bases of the Sp(4,$\mathbb{R}$) Lie algebra
by using an almost complex structure $J$, which is a skew symmetric real matrix with $J^2=-1$. 
For the Sp(4,$\mathbb{R})$ case, we take the representation
\begin{equation}
J= \left( \begin{array}{@{\,}cccc@{\,}} 0 & 1 & 0 & 0\\ -1 & 0 & 0 & 0\\  0 & 0 & 0 & 1\\ 0 & 0 & -1 & 0\\
\end{array}\right)   .
\label{3e3}
\end{equation}
Then, the Lie algebra of Sp(4,$\mathbb{R}$) is represented by the spaces of matrices $A=(a_{ij}), (1\le i, j \le 4)$, 
which satisfy $J A+A^T J=0$. It is a ten-dimensional representation spanned by the following basis elements
\begin{alignat*}{4}
I_1&=
\begin{pmatrix}
0&1&0&0\\
0&0&0&0\\
0&0&0&0\\
0&0&0&0
\end{pmatrix},
&\
I_2&=
\begin{pmatrix}
0&0&0&0\\
1&0&0&0\\
0&0&0&0\\
0&0&0&0
\end{pmatrix},
&\ 
I_3&=
\begin{pmatrix}
1&0&0&0\\
0&-1&0&0\\
0&0&0&0\\
0&0&0&0
\end{pmatrix},
&\ \ 
I_4&=
\begin{pmatrix}
0&0&1&0\\
0&0&0&0\\
0&0&0&0\\
0&-1&0&0
\end{pmatrix},
\\
I_5&=
\begin{pmatrix}
0&0&0&0\\
0&0&0&1\\
-1&0&0&0\\
0&0&0&0
\end{pmatrix},
&\
I_6&=
\begin{pmatrix}
0&0&0&1\\
0&0&0&0\\
0&1&0&0\\
0&0&0&0
\end{pmatrix},
&\
I_7&=
\begin{pmatrix}
0&0&0&0\\
0&0&1&0\\
0&0&0&0\\
1&0&0&0
\end{pmatrix},
&\
I_8&=
\begin{pmatrix}
0&0&0&0\\
0&0&0&0\\
0&0&0&1\\
0&0&0&0
\end{pmatrix},
\\
I_9&=
\begin{pmatrix}
0&0&0&0\\
0&0&0&0\\
0&0&0&0\\
0&0&1&0
\end{pmatrix},
&\
I_{10}&=
\begin{pmatrix}
0&0&0&0\\
0&0&0&0\\
0&0&1&0\\
0&0&0&-1
\end{pmatrix}.
\end{alignat*}
%
\subsection{ Lax pairs of the first and second flows}
The Lax equation of the first flow is given by
\begin{equation}
\frac{\partial L}{\partial t}=[B_2, L], 
\label{3e4}
\end{equation}
where the operators $L$ and $B_2$ have the following form
\begin{align}
L =&\frac{c_0}{2 \lambda}  
\left( \begin{array}{@{\,}cccc@{\,}} 
\gamma_1 &  \gamma_2 &  \gamma_3 &         0\\
\gamma_2 & -\gamma_1 &         0 & -\gamma_3\\
\gamma_3 &         0 & -\gamma_1 &  \gamma_2\\
       0 & -\gamma_3 &  \gamma_2 &  \gamma_1\\
\end{array}\right)
+\left( \begin{array}{@{\,}cccc@{\,}} 
       0 &         0 & -\omega_2 & -\omega_1\\ 
       0 &         0 &  \omega_1 & -\omega_2\\
\omega_2 & -\omega_1 & -2\lambda & -\omega_3\\
\omega_1 &  \omega_2 &  \omega_3 &  2\lambda\\
\end{array}\right)
\label{3e5}
\\
=&\frac{c_0}{2 \lambda} \Big(\gamma_1 (I_3-I_{10})+\gamma_2 (I_1+I_2+I_8+I_9)
+\gamma_3 (I_4-I_5)\Big)
\nonumber\\
&+\omega_1 (-I_6+I_7)+\omega_2 (-I_4-I_5)+\omega_3 (-I_8+I_9)
-2 \lambda I_{10}    ,   
\label{3e6}\\
B_2=&\frac{1}{2}  
\left( \begin{array}{@{\,}cccc@{\,}} 
        0 &  \omega_3 & -\omega_2 & -\omega_1\\ 
-\omega_3 &         0 &  \omega_1 & -\omega_2\\
 \omega_2 & -\omega_1 & -2\lambda & -\omega_3\\
 \omega_1 &  \omega_2 &  \omega_3 &  2\lambda\\
\end{array}\right)
\label{3e7}\\
=&\frac{1}{2} \omega_1 (-I_6+I_7)+\frac{1}{2} \omega_2 (-I_4-I_5)
+\frac{1}{2} \omega_3 (I_1-I_2-I_8+I_9)-\lambda I_{10} ,
\label{3e8}
\end{align}
where $\lambda$ is a constant spectral parameter. For the first flow, we denote the time-evolving 
operator as $B_2$ instead of $B_1$, because $\dd t$ is proportional to the second argument 
$\dd u_2$ of the Jacobi's inversion relationship of Eq.(\ref{2e47}). Eq.(\ref{3e4}) 
gives Eqs.(\ref{2e1})--(\ref{2e6}). In constructing a Lax pair of $L$ and $B_2$, eight Lie algebra elements
\begin{align*}
(I_1+I_2+I_8+I_9), \quad (I_1-I_2-I_8+I_9), \quad (I_3-I_{10}), 
\\
(I_4-I_5), \quad (I_4+I_5), \quad (I_6-I_7), \quad (I_8-I_9), \quad I_{10}, 
\end{align*}
emerge, i.e., $(I_1+I_9)$, $(I_2+I_8)$, $(I_6-I_7)$, $(I_8-I_9)$, $I_3$, $I_4$, $I_5$, and $I_{10}$ are necessary. 
Then, in the case of the first flow, we cannot conclude that all ten bases of Sp(4,$\mathbb{R}$) Lie algebra are necessary.

Next, we construct the Lax pair of the second flow. The operator $L$ is the same as Eq.(\ref{3e5}). 
Assuming that the second flow can be written by the basis of the Sp(4,$\mathbb{R}$) 
Lie algebra, we rearrange $B_1$ in the form
\begin{align}
B_1&=\left( \begin{array}{@{\,}rrrr@{\,}} 
 a_{11} & a_{12} & a_{13} & a_{14}\\ 
 a_{21} &-a_{11} & a_{23} & a_{24}\\
-a_{24} & a_{14} & a_{33} & a_{34}\\
 a_{23} &-a_{13} & \hphantom{-}a_{43} &-a_{33}\\
\end{array}\right)
\label{3e11}\\
&=a_{11} I_3+a_{12} I_1+ a_{13} I_4 +a_{14} I_6+a_{21} I_2
\nonumber\\
&
\quad+ a_{23} I_7 +a_{24} I_5
+a_{33} I_{10} +a_{34} I_8+a_{43} I_9   .
\label{3e12}
\end{align}
The Lax equation for the second flow is given by
\begin{equation}
\frac{\partial L}{\partial r}=[B_1, L]  .
\label{3e13}
\end{equation}
To ensure that this Lax equation gives Eqs.(\ref{2e57})--(\ref{2e62}), ten independent 
linear relationships for ten variables 
$a_{11}$, $a_{12}$, $a_{13}$, $a_{14}$, $a_{21}$, $a_{23}$, $a_{24}$, $a_{33}$, $a_{34}$, and $a_{43}$ must 
be satisfied. These ten equations to be satisfied are provided in 
Appendix A. However, the expressions of the solutions are 
too long, and the entire expressions cannot be presented in this paper; hence, 
we present only the first few terms in Appendix A. To achieve our , 
it is not necessary to know the explicit form 
of  $B_1$; however it is sufficient that there really exist solutions. 
By using these solutions $a_{ij}$'s, we have confirmed that the Lax equation 
$\partial_r L=[B_1, L]$ is completely satisfied by using the second flow 
equations Eqs.(\ref{2e57})--(\ref{2e62}). As $a_{23}=-1, a_{24}=-2$, we 
have $a_{24}=2a_{23}$; however other $a_{ij}$'s are independent, 
such that we need nine bases 
\begin{equation}
I_1,  \quad I_2,  \quad I_3, \quad I_4, \quad I_6, \quad I_8, \quad I_9, \quad I_{10}, \quad (I_7+2 I_5),
\label{3e26}
\end{equation}
for the Lax pair of the second flow. {However, for the first flow, $I_5$ is required. 
Then $I_7$ itself becomes necessary in the combination of both flows. This fact implies 
that all of the ten \mbox{Sp(4,$\mathbb{R}$)/$\Z_2$} $\cong$ SO(3,2) Lie algebra bases 
are necessary and sufficient to construct Lax pairs of the first and second flows 
for two-flows Kowalevski top. It is important that coefficients of all ten Sp(4,$\mathbb{R}$)/$\Z_2$ Lie 
algebra bases are independent. If some coefficients of the  Lie algebra 
bases are not independent, there is a possibility that the structure of the Lie algebra of two 
flow Kowalevski top reduces to some Lie subalgebra of the Sp(4,$\mathbb{R}$)/$\Z_2$ Lie algebra. In the 
Lax formalism, it is quite non-trivial and is quite difficult to  find the necessary and 
sufficient Lie algebra to obtain the two-flows Kowalevski top. In such a situation, we say 
that the two-flows Kowalevski top has the Lie group structure.

On one hand, we demonstrated that the two-flows Kowalevski 
top is equivalent to the full genus two Jacobi's inversion problem in Section 2. 
On the other hand, we proved that the Lie group structure of Lax pairs of 
two-flows Kowalevski top is Sp(4,$\mathbb{R}$)/$\Z_2$ $\cong$ SO(3,2) in Section 3. 
Combining these results, we conclude that the genus two hyperelliptic function has 
the Sp(4,$\mathbb{R}$)/$\Z_2$ $\cong$ SO(3,2) structure.

\section{Summary and Discussions} 
\setcounter{equation}{0}

First, we reviewed the Kowalevski top to explain the formulation and  
notation, as it is quite prevalent, yet was first introduced more than one hundred years ago. 
The equations of the original Kowalevski top, or the first flow, can be formulated into the special genus 
two Jacobi's inversion problem. In addition to the first flow, we have adopted the second 
flow, i.e., other equations of the Kowalevski top by introducing another time $r$. The first 
and second flows satisfy the integrability condition. In addition, 
we demonstrated  that the 
two-flows Kowalevski top is formulated into the full genus two Jacobi's inversion problem.

Next, in addition to the Lax pair of the first flow for the Kowalevski top, we  
constructed the Lax pair of the second flow for the Kowalevski top. Then, we 
deduced the Sp(4,$\mathbb{R}$)/$\Z_2$ $\cong$ SO(3,2) 
Lie group structure of Lax pairs of the first and second flows. 
Combining these results, using the two-flows Kowalevski top, we 
infer that the genus two hyperelliptic function, 
which is the solution of the full genus two Jacobi's inversion problem has 
the Sp(4,$\mathbb{R}$)/$\Z_2$  $\cong$ SO(3,2) Lie group structure.

The Lie group structure is Sp(2,$\mathbb{R}$) for the genus one elliptic function 
and Sp(4,$\mathbb{R}$) for the genus two hyperelliptic function, 
which resembles the Siegel's discrete Sp($2g$, $\mathbb{Z}$) modular transformation 
for the genus $g$ hyperelliptic theta functions~\cite{Siegel}.

\appendix
\section{Ten equations for $\bm{a_{ij}}$'s in the second flow Lax equations and expression of the solution}
\setcounter{equation}{0}
Ten equations for $a_{ij}$'s in the second flow Lax equations
are given in the form
\begin{align}
1)&\quad a_{12} \gamma_2 c_0+a_{13}(\gamma_3 c_0+2 \omega_2 \lambda)
+2 a_{14}\omega_1 \lambda -a_{21} \gamma_2 c_0 +2 a_{23}\omega_1 \lambda
+a_{24} (\omega_3 c_0-2\omega_2\lambda)\nonumber\\
&\quad -\gamma_3 c_0(\gamma_1 \omega_2-\gamma_2 \omega_1)
+\gamma_3 \omega_3(\omega_1^2+\omega_2^2)=0,
\label{A1}\\
2)&\quad 2 a_{11} \gamma_2 c_0 -2 a_{12} \gamma_1 c_0-4 a_{13}\omega_1 \lambda 
-2 a_{14}(\gamma_3 c_0-2 \omega_2 \lambda)
-\gamma_3 c_0(\gamma_1 \omega_1+\gamma_2 \omega_2) \nonumber\\
&\quad -\gamma_3 \omega_1(\omega_1^2+\omega_2^2)=0,
\label{A2}\\
3)&\quad a_{11} c_0(\gamma_3 c_0-2 \omega_2 \lambda)
+2 a_{12} \omega_1 c_0 \lambda -2 a_{13}c_0 (\gamma_1 c_0+2\lambda^2)
+a_{14}c_0(\gamma_2 c_0+2 \omega_3 \lambda)  \nonumber\\
&\quad - a_{23}\gamma_2 c_0^2 -a_{33} c_0( \omega_3 c_0-2\omega_2\lambda)
+2 a_{43} \omega_1 c_0 \lambda 
+ \omega_2 c_0^2(\gamma_1^2+ \gamma_2^2)
+\omega_3 c_0 \lambda (\gamma_1 \omega_1+\gamma_2 \omega_2)\nonumber\\
&\quad -(\gamma_1 \omega_2 c_0-\gamma_2 \omega_1 c_0-\omega_1 \omega_3 \lambda)
(\omega_1^2+\omega_2^2) 
+\gamma_3 c_0 \lambda (\gamma_1 c_0 +\omega_1^2-\omega_2^2)
=0,
\label{A3}\\
4)&\quad -2 a_{11} \omega_1 c_0 \lambda -a_{12} c_0(\gamma_3 c_0+2 \omega_2 \lambda)
+a_{13} c_0 (\gamma_2 c_0-2 \omega_3 \lambda)+4 a_{14} c_0 \lambda^2
-a_{24} \gamma_2 c_0^2 \nonumber\\
&\quad -2 a_{33}\omega_1 c_0 \lambda -a_{34}c_0(\gamma_3 c_0-2 \omega_2 \lambda)
+\omega_3 c_0 \lambda(\gamma_1 \omega_2-\gamma_2 \omega_1)
-\gamma_3 c_0 \lambda (\gamma_2 c_0+2\omega_1 \omega_2) \nonumber\\
&\quad -\omega_2 \omega_3 \lambda (\omega_1^2+\omega_2^2)=0,
\label{A4}\\
5)&\quad -2 a_{11} \gamma_2 c_0 +2 a_{21} \gamma_1 c_0
+2 a_{23}( \omega_3 c_0+2\omega_2\lambda) +4 a_{24}\omega_1 \lambda
-\gamma_3 c_0(\gamma_1 \omega_1+\gamma_2 \omega_2) \nonumber\\
&\quad -\gamma_3 \omega_1(\omega_1^2+\omega_2^2)=0,
\label{A5}\\
6)&\quad -2 a_{11} \omega_1 c_0 \lambda -a_{13} \gamma_2 c_0^2
+a_{21} c_0(\gamma_3 c_0-2 \omega_2 \lambda)-4a_{23} c_0 \lambda^2
+a_{24} c_0(\gamma_2 c_0+2 \omega_3 \lambda) \nonumber\\
&\quad -2 a_{33}\omega_1 c_0 \lambda 
+a_{43}c_0(\gamma_3 c_0+2 \omega_2 \lambda)
-\omega_3 c_0 \lambda(\gamma_1 \omega_2-\gamma_2 \omega_1)
+\gamma_3 c_0 \lambda (\gamma_2 c_0+2\omega_1 \omega_2)\nonumber\\
&\quad +\omega_2 \omega_3 \lambda (\omega_1^2+\omega_2^2)=0,
\label{A6}\\
7)&\quad a_{11} c_0(\gamma_3 c_0+2 \omega_2 \lambda)
-a_{14} \gamma_2 c_0^2 -2 a_{21} \omega_1 c_0 \lambda 
+a_{23}c_0(\gamma_2 c_0-2 \omega_3 \lambda) +a_{24}(\gamma_1 c_0+2 c_0 \lambda) 
\nonumber\\
&\quad -a_{33} c_0( \omega_3 c_0+2\omega_2\lambda)
-2 a_{34} \omega_1 c_0 \lambda
- \omega_2 c_0^2(\gamma_1^2+ \gamma_2^2)
+\omega_3 c_0 \lambda (\gamma_1 \omega_1+\gamma_2 \omega_2)\nonumber\\
&\quad +(\gamma_1 \omega_2 c_0-\gamma_2 \omega_1 c_0+\omega_1 \omega_3 \lambda)
(\omega_1^2+\omega_2^2)
+\gamma_3 c_0 \lambda (\gamma_1 c_0 +\omega_1^2-\omega_2^2)=0,
\label{A7}\\
8)&\quad -a_{13}( \omega_3 c_0+2\omega_2\lambda) +2a_{14}\omega_1 \lambda
+2a_{23}\omega_1 \lambda -a_{24}( \omega_3 c_0-2\omega_2\lambda) 
+a_{34}( \gamma_2 c_0+2\omega_3\lambda)        \nonumber\\
&\quad  -a_{43}( \gamma_2 c_0-2\omega_3\lambda)
+\gamma_3 c_0(\gamma_1 \omega_2-\gamma_2 \omega_1) 
-\gamma_3 \omega_2(\omega_1^2+\omega_2^2)=0,
\label{A8}\\
9)&\quad -2a_{14} (\gamma_3 c_0+2 \omega_2 \lambda)
+4 a_{24}\omega_1 \lambda +2a_{33} (\gamma_2 c_0-2 \omega_3 \lambda)
+2 a_{34}(\gamma_1 c_0+4\lambda^2)  \nonumber\\
&\quad -(\gamma_3 c_0+2 \omega_2 \lambda)(\gamma_1 \omega_1+\gamma_2 \omega_2)
-2\omega_1 \lambda (\gamma_1 \omega_2-\gamma_2 \omega_1)
-\gamma_3 \omega_1(\omega_1^2+\omega_2^2)=0,
\label{A9}\\
10)&\quad -4a_{13}\omega_1 \lambda  +2a_{23} (\gamma_3 c_0-2 \omega_2 \lambda)
-2a_{33} (\gamma_2 c_0+2 \omega_3 \lambda)
-2 a_{43}(\gamma_1 c_0+4\lambda^2)  \nonumber\\
&\quad -(\gamma_3 c_0-2 \omega_2 \lambda)(\gamma_1 \omega_1+\gamma_2 \omega_2)
+2\omega_1 \lambda (\gamma_1 \omega_2-\gamma_2 \omega_1)
-\gamma_3 \omega_1(\omega_1^2+\omega_2^2)=0. 
\label{A10}
\end{align}
The first few terms of $a_{ij}$'s are given by
\begin{align}
1)&\quad  a_{11}=\frac{1}{2c_0 D}(16\gamma_1^5 \gamma_3 \omega_1^2 c_0^5
+32\gamma_1^5 \omega_1^2 \omega_2  c_0^4 \lambda
+32\gamma_1^4 \gamma_2 \gamma_3 \omega_1 \omega_2  c_0^5 +\cdots)  , 
\label{3e14}\\
2)&\quad  a_{12}=\frac{1}{2c_0 D}(16\gamma_1^4 \gamma_2 \gamma_3 \omega_1^2 c_0^5
+32\gamma_1^4 \gamma_2 \omega_1^2 \omega_2  c_0^4 \lambda
+32\gamma_1^3 \gamma_2^2 \gamma_3 \omega_1 \omega_2  c_0^5 +\cdots)  , 
\label{3e15}\\
3)&\quad  a_{13}=\frac{1}{D}(8\gamma_1^4 \gamma_3^2 \omega_1^2 c_0^4
-32\gamma_1^4 \omega_1^2 \omega_2^2  c_0^2 \lambda^2
+16\gamma_1^3 \gamma_2 \gamma_3^2 \omega_1 \omega_2  c_0^4 +\cdots)  ,
\label{3e16}\\
4)&\quad a_{14}=\frac{1}{D}
(-32\gamma_1^3 \gamma_3 \omega_1^2 \omega_2 \omega_3 c_0^2 \lambda^2
-64\gamma_1^3 \omega_1^2 \omega_2^2 \omega_3 c_0 \lambda^3
+32\gamma_1^2 \gamma_2 \gamma_3^2 \omega_1^2 c_0^3 \lambda^2 +\cdots)  , 
\label{3e17}\\
5)&\quad a_{21}=\frac{1}{2c_0 D}(16\gamma_1^4 \gamma_2 \gamma_3 \omega_1^2 c_0^5
+32\gamma_1^4 \gamma_2 \omega_1^2 \omega_2  c_0^4 \lambda
+32\gamma_1^3 \gamma_2^2 \gamma_3 \omega_1 \omega_2  c_0^5 +\cdots)  ,  
\label{3e18}\\
6)&\quad a_{23}=-1 , 
\label{3e19}\\
7)&\quad a_{24}=-2  , 
\label{3e20}\\
8)&\quad a_{33}=\frac{1}{2c_0 D}(-16\gamma_1^5 \gamma_3 \omega_1^2 c_0^5
-32\gamma_1^5 \omega_1^2 \omega_2  c_0^4 \lambda
-32\gamma_1^4 \gamma_2 \gamma_3 \omega_1 \omega_2  c_0^5 +\cdots)  , 
\label{3e21}\\
9)&\quad a_{34}=\frac{1}{2c_0 D}(16\gamma_1^4 \gamma_2 \gamma_3 \omega_1^2 c_0^5
+32\gamma_1^4 \gamma_2 \omega_1^2 \omega_2  c_0^4 \lambda
-32\gamma_1^4 \gamma_3 \omega_1^2 \omega_3  c_0^4 \lambda +\cdots)  , 
\label{3e22}\\
10)&\quad a_{43}=\frac{1}{2c_0 D}(16\gamma_1^4 \gamma_2 \gamma_3 \omega_1^2 c_0^5
+32\gamma_1^4 \gamma_2 \omega_1^2 \omega_2  c_0^4 \lambda
+32\gamma_1^4 \gamma_3 \omega_1^2 \omega_3  c_0^4 \lambda +\cdots)  , 
\label{3e23}
\end{align}
with
\begin{equation}
D=4\gamma_1^4 \gamma_3^2 \omega_1^2 c_0^4
+16 \gamma_1^4 \gamma_3 \omega_1^2 \omega_2  c_0^3 \lambda
+16\gamma_1^4 \omega_1^2 \omega_2^2  c_0^2 \lambda^2 +\cdots  .
\label{3e24}
\end{equation}
It appears that $a_{12}=a_{21}$ and $a_{33}=-a_{11}$ hold at first glance; however, 
they are actually different. 
Namely, $a_{12}\ne a_{21}$ and $a_{33}\ne-a_{11}$.

\newpage

\end{document}